\documentclass[aps,prl,groupedaddress,twocolumn]{revtex4-2}
\usepackage{amssymb,amsmath}
\usepackage{graphicx}
\usepackage{graphics}
\usepackage{epsfig}
\usepackage{color}
\newcommand{\fig}[1]{\parbox{1.5cm}{\epsfig{file=#1.eps,width=1.5cm}}}

\begin{document}

\title{Casimir contribution to the interfacial Hamiltonian for 3D wetting}

\author{Alessio Squarcini}
\affiliation{Max Planck Institute for Intelligent Systems, Heisenbergstr. 3, D-70569 Stuttgart, Germany \vspace{1mm}}
\affiliation{IV. Institut f\"ur Theoretische Physik, Universit\"at Stuttgart, Pfaffenwaldring 57, D-70569 Stuttgart, Germany \vspace{1mm}}
\affiliation{Institut f\"ur Theoretische Physik, Universit\"at Innsbruck, Technikerstra\ss e 21A, A-6020, Innsbruck, Austria \vspace{1mm}}
\author{Jos\'e M. Romero-Enrique}
\affiliation{Departamento de F\'{\i}sica At\'omica, Molecular y Nuclear, \'Area de F\'{\i}sica Te\'orica, Universidad de Sevilla, Avenida de Reina Mercedes s/n, 41012 Seville, Spain}
\author{Andrew O. Parry}
\affiliation{
Department of Mathematics, Imperial College London, London SW7 2AZ, United Kingdom}

\date{\today}

\begin{abstract}
Previous treatments of three-dimensional (3D) short-ranged wetting transitions have missed an entropic or low temperature Casimir contribution to the binding potential describing the interaction between the unbinding interface and wall. This we determine by exactly deriving the interfacial model for 3D wetting from a more microscopic Landau-Ginzburg-Wilson Hamiltonian. The Casimir term changes the interpretation of fluctuation effects occurring at wetting transitions so that, for example, mean-field predictions are no longer obtained when interfacial fluctuations are ignored. While the Casimir contribution does not alter the surface phase diagram, it significantly increases the adsorption near a first-order wetting transition and changes completely the predicted critical singularities of tricritical wetting, including the non-universality occurring in 3D arising from interfacial fluctuations. Using the numerical renormalization group we show that, for critical wetting, the asymptotic regime is extremely narrow with the growth of the parallel correlation length characterised by an effective exponent in quantitative agreement with Ising model simulations, resolving a longstanding controversy.
\end{abstract}

\maketitle

Interfaces between solids and fluids exhibit a wealth of physics which can often be studied using effective models motivated by mesoscopic principles e.g. the capillary-wave model of interfacial wandering \cite{buff} and models of surface growth \cite{KPZ}. However, for critical wetting in 3D systems with short-ranged forces the details of the interfacial model, and understanding how it emerges from a microscopic framework, is of crucial importance. Critical wetting refers to the continuous growth of a liquid phase (for example) at a solid-gas interface (wall) as the temperature is increased towards a wetting temperature and is associated with the divergence of a parallel correlation length, $\xi_\parallel$, characterised by an exponent $\nu_\parallel$. For comprehensive reviews see \cite{dietrich_wetting_1988,Schick,FLN,BEIMR}. In general, this can be described accurately using a simple interfacial model incorporating the surface tension (or stiffness) and a bind-potential determined by integrating the intermolecular forces over the volume of liquid. Greater care is required in 3D with short-ranged forces where the binding potential itself arises from density fluctuations and decays on the scale of the bulk correlation length. 3D is the upper critical dimension and the original renormalization group (RG) studies predicted strong non-universal critical singularities \cite{LKZ_1983,BHL_1983,FH_1985} implying that $\nu_\parallel \approx 3.7$ for Ising-like systems, very different to the mean-field prediction $\nu_\parallel=1$. However, these have never been seen in experiments \cite{BONN_NATURE, BONN_PRL} nor very careful Ising model simulations \cite{BLK_1986,BL_1988,PEB_1991, BLW_1989, BB_2013, BT_2021} which observe instead an effective exponent $\nu_{\parallel}^{{\rm eff}} = 1.8 \pm 0.1$. Allowing for a position dependence to the stiffness exacerbated the problem since the transition is driven first-order \cite{FJ_1992, JF_1993, Boulter_1997} -- a prediction not seen in the simulations and which also contradicts the expected Nakanishi-Fisher global surface phase diagrams which connects consistently wetting to surface criticality \cite{NF}. A likely factor in the resolution of this controversy is that the binding potential is, in general, non-local arising from correlations within the wetting layer \cite{PREL_2004, PRBRE_2008_prl, PRBRE_2006, RESPG_2018}. This can be expressed using a compact diagrammatic formulation which can be used for walls of arbitrary shape. In this way, the possibility that the wetting transition is driven first-order is removed completely so that the global phase diagram is restored.

While this progress is encouraging there is a problem. Since wetting occurs below the critical temperature, $T_c$, it has been assumed that bulk-like fluctuations are unimportant, and it is only the (thermal) wandering of the interface that leads to non-classical exponents. This is explicit in derivations of interfacial Hamiltonians from more microscopic Landau-Ginzburg-Wilson (LGW) models in which they are identified via a constrained minimization, equivalent to a mean-field (MF) approximation of the trace over microscopic degrees of freedom. This means that binding potentials have been missing an entropic contribution arising from the multiplicity of microscopic configurations that correspond to a given interfacial one -- a feature which is known to be important in molecular descriptions of free interfaces \cite{Pedro_1, Pedro_2, Pedro_3,MCT}. The entropic contribution to the binding potential is akin to a thermal Casimir effect -- the force between two walls due to the restriction of bulk fluctuations in a confined fluid \cite{FdG, BCN, ES, HGDB}. At $T_c$ this force is long-ranged but it is always present, even away from the critical point where it decays on the scale of the bulk correlation length \cite{AM2010,DS2015}. For short-ranged wetting there is, therefore, an additional entropic or low-temperature Casimir term in the binding potential, which is a similar range to the MF contribution. In this paper we determine this using the non-local, diagrammatic, formalism by performing properly the constrained trace for the LGW model, which exactly determines the interfacial Hamiltonian for 3D wetting. We show that the Casimir term plays an important role at wetting transitions of all orders forcing a reappraisal of the accuracy of MF and subsequent RG theory and altering even the values of critical exponents.

Our starting point is the LGW Hamiltonian based on a magnetization-like order-parameter 
(see \cite{NF})
\begin{equation}
H[m]= \int d\mathbf{r} \left(\frac{1}{2} (\boldsymbol\nabla m)^2 +\phi(m)\right) + \int_{\mathcal{S}_\psi} d\mathbf{s}  \phi_1 (m(\mathbf{s})),
\label{HLGW}
\end{equation}
where $\phi(m)$ is a double well potential which we assume has an Ising symmetry and denote $m_0$ the spontaneous magnetization and $\kappa$ the inverse bulk correlation length. Here $\phi_1= -g(m-m_s)^2/2$ is the surface potential with $g$ the enhancement parameter and $m_s$ the favoured order-parameter at the wall $\mathcal{S}_\psi$ with Monge parameterization $(\mathbf{x}, \psi)$. Equivalently $h_1=-gm_s$ is the surface field. Minimizing $H[m]$ determines the MF phase diagram which for a planar wall shows critical wetting (when $-g>\kappa$) and first-order wetting transitions (for $-g<\kappa$). We can also use it 
to derive an interfacial model $H_I[\ell]$ with the interfacial co-ordinate determined by a crossing criterion so that $m(\mathbf{x}, \ell(x))=0$ on the interface $\mathcal{S}_\ell$. Formally, this is identified via $\exp(-\beta H_I[\ell])=\int \mathcal{D}'m \exp (-\beta H[m])$ where $\beta=1/k_BT$ and the prime denotes a constrained trace over microscopic 
degrees of freedom respecting the crossing criterion \cite{FJ_1991}. This yields
\begin{equation}
\label{ }
H_I[\ell]=\gamma A_{l}+W[\ell,\psi] \, ,
\end{equation}
where the first term is the surface tension times the interfacial area describing the free interface (ignoring curvature 
terms) and $W[\ell,\psi]$ is the binding potential functional describing the interaction with the wall. 
 
To evaluate the constrained trace it is now customary to ignore bulk fluctuations and make a MF approximation which identifies $H_I[\ell]=H[m_\Xi]$ where $m_\Xi$ is the \emph{unique} profile that 
minimizes the LGW Hamiltonian subject to the 
crossing criterion. Within the reliable double parabola (DP)
approximation $\phi(m)=\kappa^2(|m|- m_0)^2/2$ this can 
be done analytically. For example, 
if the interface is a uniform 
thickness $\ell$ from a planar wall ($\psi=0$), of lateral 
area $L_\parallel^2$, then the binding potential functional reduces to the binding potential function $w_{MF}=W_{MF}/ L_\parallel^2$ which has the well-known exponential expansion \cite{FJ_1991}
\begin{equation}
\frac{w_{MF}(\ell)}{\gamma}\approx -\frac{2t  g }{g-\kappa} \textrm{e}^{-\kappa \ell} +\frac{g+\kappa}{g-\kappa} \textrm{e}^{-2\kappa \ell},
\label{wflatmf}
\end{equation}
where $\gamma=\kappa m_0^2$ is the surface tension 
and $t=(m_0-m_s)/m_0$ is the temperature-like scaling 
field for 
critical wetting. 
The minimum of $w_{MF}$ determines the MF wetting layer thickness while its curvature at this point determines $\xi_\parallel$. Both these lengthscales diverge continuously as $t\to 0$ when $-g>\kappa$. Note that
for tricritical wetting ($g=-\kappa$) and first-order wetting ($-g<\kappa$) it is necessary to include the next-order 
decaying exponential term. For non-planar interfaces 
(and walls) the non-local MF functional $W_{MF}[\ell,\psi]$ can also be determined exactly using boundary integral methods based on the Green function in the wetting layer \cite{PRBRE_2006,RESPG_2018}. However, within the DP approximation, the constrained trace can be performed determining the exact binding potential functional
\begin{equation}
\label{ }
W[\ell,\psi]=W_{MF}[\ell,\psi] +W_{C}[\ell,\psi] \, ,
\end{equation}
which contains a Casimir correction. The proof of this is rather technical but is outlined in the Supplementary Material \cite{SM}. Full details will be published elsewhere \cite{jointpublication}. Here we limit ourselves to the final result for $W_C[\ell,\psi]$ and the implications for wetting transitions of all orders. The Casimir contribution can be represented diagrammatically similar to the terms in the mean-field contribution but has a distinct topology. To this end we introduce two kernels which connect positions, with respective transverse co-ordinates $\textbf{s}$, $\textbf{s}^{\prime}$ (denoted by the open circles) on the interface (upper wavy line) and wall (lower wavy line). We have
\begin{equation}
\fig{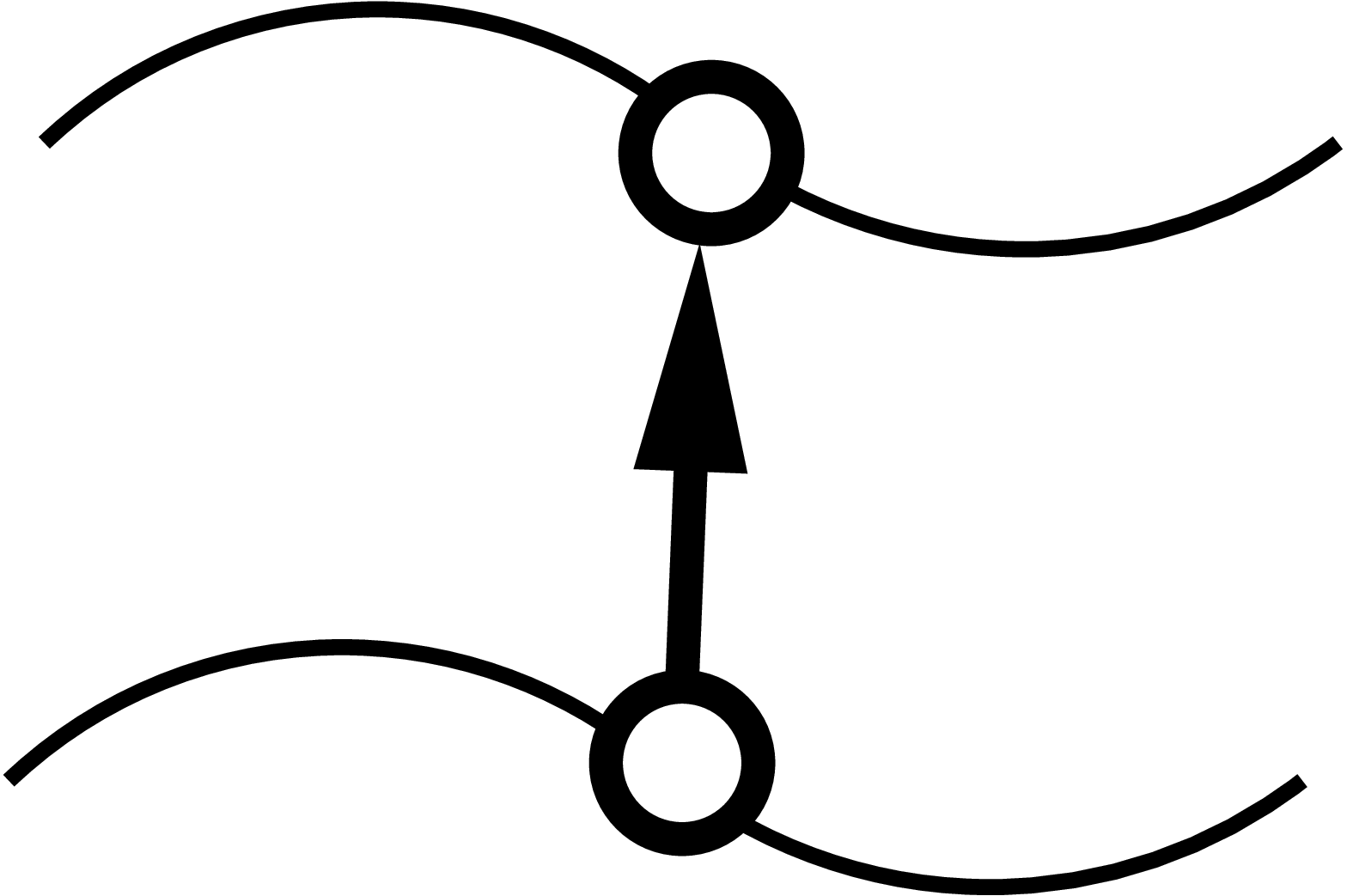} =  \mathbf{n}(\mathbf{s}) \mathbf{\cdot} \frac{\mathbf{s}^{\prime}-\mathbf{s}}{|\mathbf{s}-\mathbf{s}^{\prime}|^2}\left(1+ \frac{1}{\kappa|\mathbf{s}-\mathbf{s}^{\prime}|}\right)\textrm{e}^{-\kappa|\mathbf{s}-\mathbf{s}^{\prime}|} \, ,
\label{defdiagrams3bisbis}
\end{equation}
which was introduced in \cite{RESPG_2018} in the derivation of $W_{MF}[\ell,\psi]$. Here $\textbf{n}(\textbf{s})$ is the normal at the wall. We also define
\begin{eqnarray}
\fig{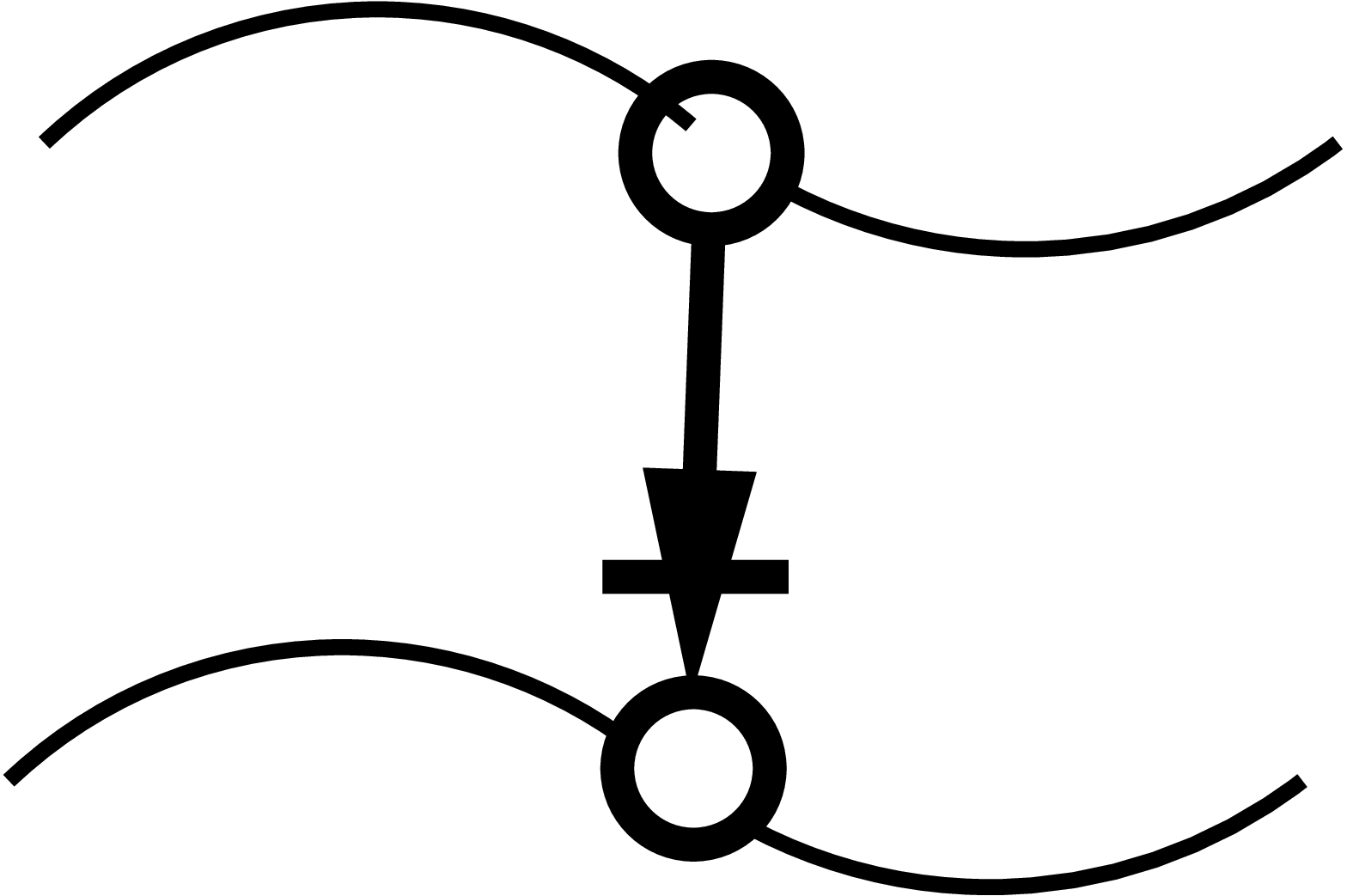} 
= \frac{1}{2\pi} \int_0^\infty dq\ q \frac{g+\kappa_q}{g-\kappa_q} J_0(q\rho)\exp(-\kappa_q \ell) \, ,
\label{defdiagrams2bisbis}
\end{eqnarray}
where $\rho$ and $\ell$ are, respectively, the transverse and normal coordinates of $\mathbf{s}-\mathbf{s}'$, $J_0(z)$ is the Bessel function of the first kind and zero order and $\kappa_q\equiv \sqrt{\kappa^2+q^2}$. In terms of these the Casimir term for a wetting film at a wall of arbitrary shape can be written
\begin{equation}
\beta W_{C}[\ell,\psi]=\frac{1}{2} \fig{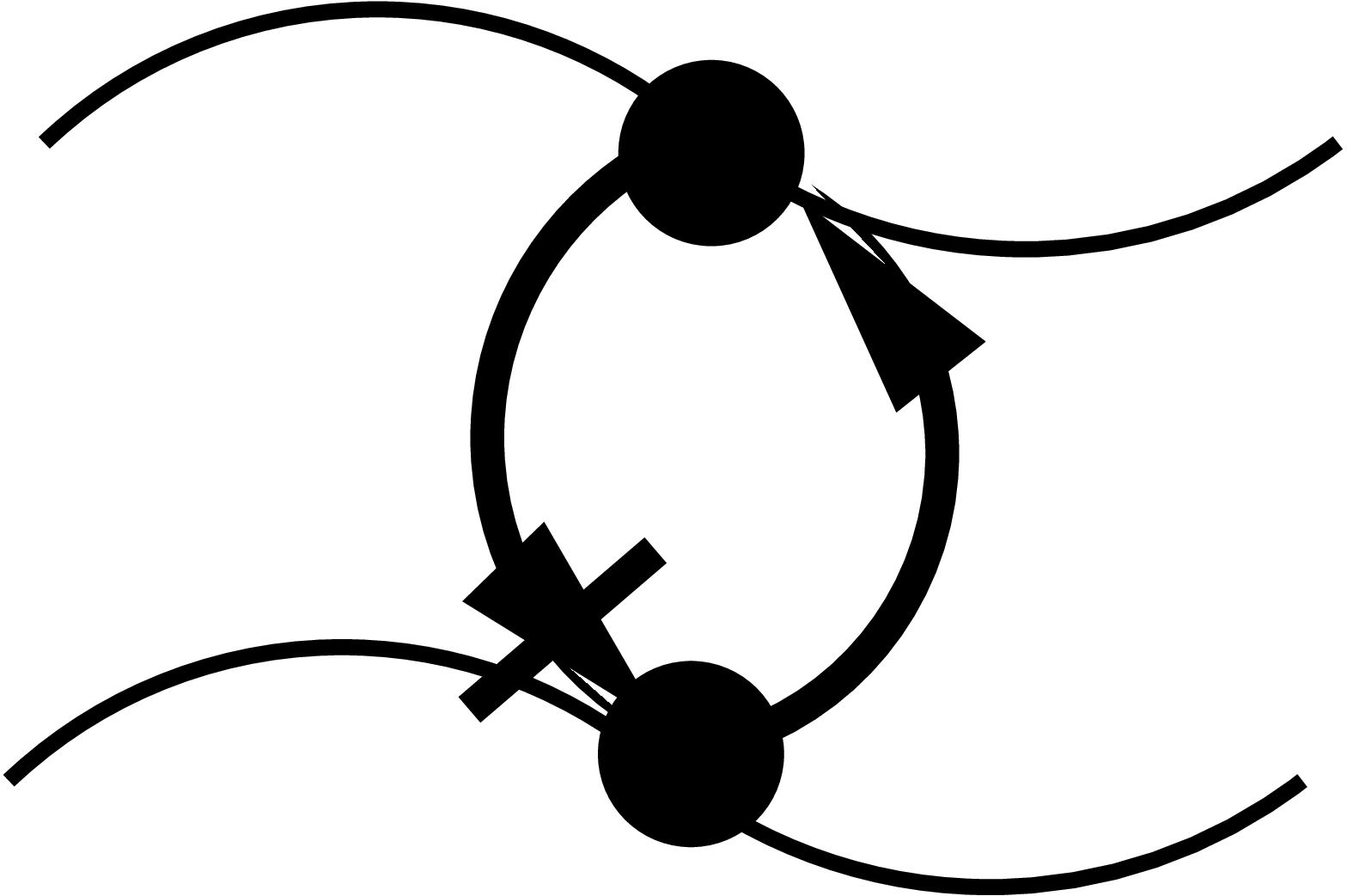} - \frac{1}{4} \fig{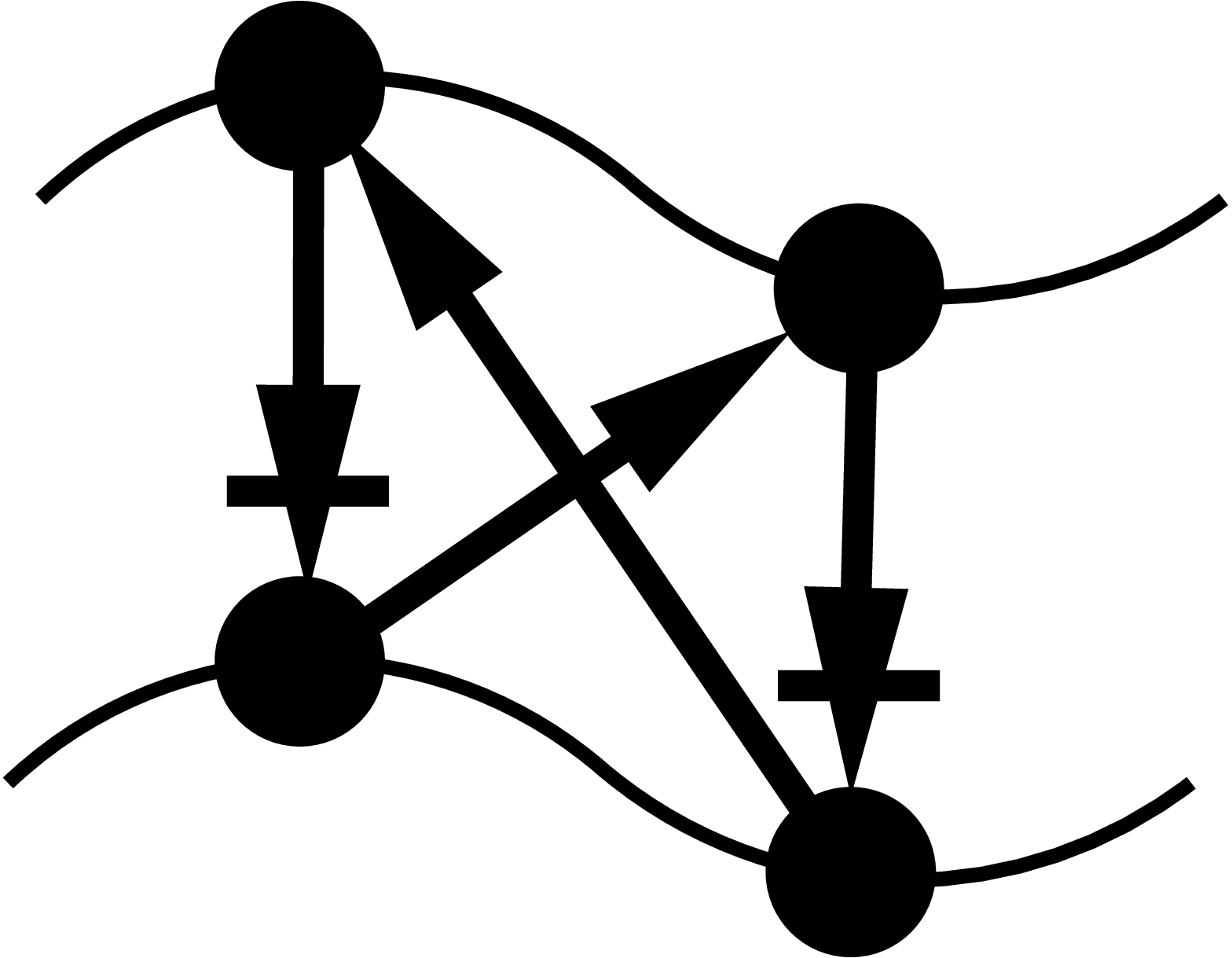}+\ldots \, ,
\label{wcasimirdiagrammatic}
\end{equation} 
which is our central result. Here the black dots simply imply integration over the points on the wall and interface with the appropriate measure for the local area. For thick wetting films, only the first term, which we refer to as $(\Omega_C)_1^1$ (see Supplementary Material \cite{SM}), is required.

\begin{figure}[t!]
\includegraphics[width=8.6cm]{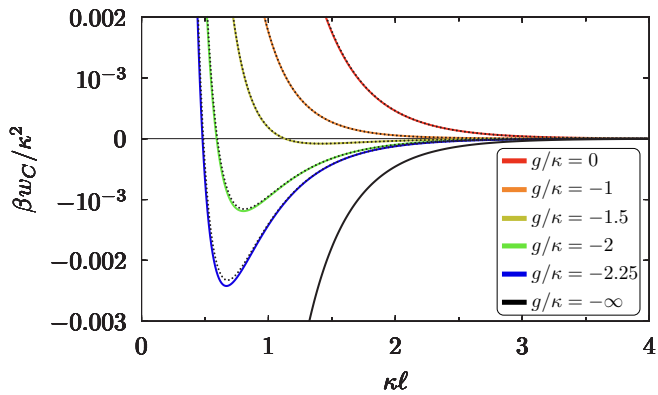}
\caption{
The Casimir contribution to the binding potential for a wetting layer of uniform thickness, for increasing surface enhancements, showing the qualitative change from attraction to repulsion near the MF tricritical point. The dotted lines show the comparisons between the full result, Eq. (\ref{wflat}), and the leading term arising from $(\Omega_C)_1^1$, which is near exact.
\label{fig6}}
\end{figure}
 
We now focus on wetting at planar walls. Before considering the 
role of interfacial fluctuations we must first 
check if the MF predictions are 
altered by determining the Casimir binding potential function $w_C(\ell)=W_C/L_\parallel^2$ for a uniform wetting layer, which adds to (\ref{wflatmf}). This can be determined exactly as \cite{jointpublication}
\begin{equation}
\beta w_{C}(\ell)= \frac{1}{4\pi}\int dq\ q\ln\left(1-\frac{g+\kappa_q}{g-\kappa_q} \textrm{e}^{-2\kappa_q\ell}\right).
\label{wflat}
\end{equation}
This is similar in form to $w_{MF}(\ell)$ but controlled by $g$ rather than $t$. For $\kappa\ell \ll 1$, $w_{C}(\ell) \propto 1/\ell^{2}$, which is the familiar long-ranged Casimir limit \cite{FdG}. More generally, for $-g>\kappa$ the potential is repulsive at short-distances, attractive at large distances and possesses a minimum which diverges continuously as $-g$ approaches  
$\kappa$. For $-g <\kappa$, the potential is purely 
repulsive (see Fig.~\ref{fig6}). In the vicinity of the MF 
tricritical point, $-g \approx \kappa$, where this
qualitative change 
occurs, the Casimir potential
behaves as
\begin{equation}
\beta w_{C}(\ell)\approx \frac{\textrm{e}^{-2\kappa \ell}}{32\pi\ell^2}(1+2(\kappa+g)\ell) \, ,
\label{asymptomegac}
\end{equation}
which is of a similar range to $w_{MF}(\ell)$. While some aspects of wetting are unchanged, others are altered completely, even before we consider the role of interfacial fluctuations. The Nakanishi-Fisher surface phase diagram 
is unaffected qualitatively so that, for example, critical wetting still occurs for $-g>\kappa$ as $t\to 0$ with $\xi_\parallel\sim t^{-1}$, as before. However, there are two significant implications. Firstly, the tricritical wetting transition is very different since it is the Casimir term that determines the repulsion. In dimension $d>1$ this decays as $\textrm{e}^{-2\kappa \ell}/\ell^{\frac{d+1}{2}}$ which, at finite $T$, {\it{always}} dominates over the higher-order MF contribution. Thus, in dimension $d>3$, where interfacial fluctuations are irrelevant, it follows that the parallel correlation length diverges as $\xi_\parallel \sim 1/ (t |\ln t| ^{\frac{d-1}{2}})$ in contrast to the strict MF prediction $\xi_\parallel \sim 1/ t^{\frac{3}{4}}$ which misses the Casimir term. MF is only recovered on setting $T=0$ or $d=\infty$. There are also consequences for first-order wetting and, in particular, the value of the film thickness $\ell_{eq}$ at the transition which, recall, smoothly increases as we follow the line of wetting transitions toward the tricritical point. MF theory predicts $\kappa \ell_{eq} \approx - \ln(1+g/\kappa)$ while, in 3D, the Casimir contribution alters this to
\begin{equation}
\ell_{eq}\approx  \frac {1} {\sqrt{16\pi \beta\gamma\left(1+\frac{g}{\kappa} \right)}}.
\end{equation}
The Casimir repulsion therefore dramatically increases the adsorption for weakly first-order transitions and similarly enhances the parallel correlation length.

Finally, we consider the non-universality occurring in 3D arising from interfacial fluctuations, controlled by the wetting parameter $\omega= k_BT \kappa^2/4\pi\Sigma$, where $\Sigma$ is the stiffness. Critical exponents for critical wetting are unchanged since $W_C$ is higher-order than $W_{MF}$. Thus we anticipate that, in the asymptotic critical regime,  
$\xi_\parallel\sim t^{-\nu_\parallel}$, where $\nu_\parallel=1/(1-\omega)$ for $0<\omega<1/2$, $\nu_\parallel=1/(\sqrt{2}-\sqrt{\omega})^2$ for $1/2<\omega<2$ and $\nu_\parallel=\infty$ for $\omega>2$ \cite{BHL_1983,FH_1985}. For tricritical wetting however the non-universality is different to previous predictions \cite{BL_1988, BC_2001} and is very similar to critical wetting containing logarithmic corrections. For
$0<\omega<1/2$, the equilibrium film thickness and parallel correlation length diverge as
\begin{equation}
\label{new_16}
\kappa \langle \ell \rangle \approx (1+2\omega)\ln\xi_\parallel - \ln\ln\xi_\parallel
\end{equation}
and
\begin{equation}
\xi_\parallel\sim (t|\ln t|)^{-\nu_\parallel},
\label{xi_parallel_RG}
\end{equation}
respectively, with $\nu_\parallel=1/(1-\omega)$. Note that on setting $\omega=0$, corresponding to infinite stiffness, these do not recover MF theory, as has been always assumed previously, but rather, the corrected results based on minimising the total binding potential $w(\ell)$ allowing for the Casimir contribution.
For  $\omega>1/2$, where the tricritical phase boundary is shifted away from $-g=\kappa$, the exponents are identical to those for critical wetting. 

\begin{figure}[t!]
\includegraphics[width=8.6cm]{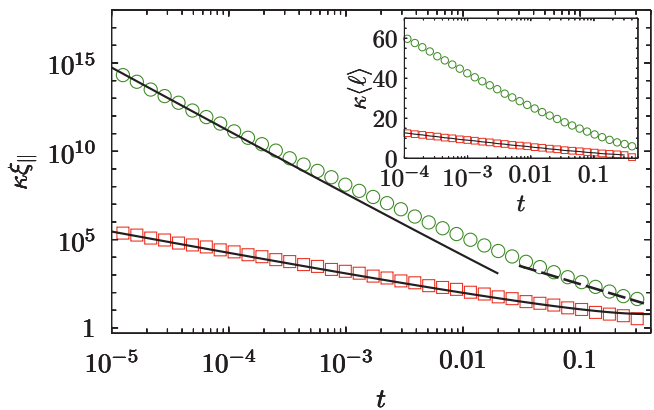}
\caption{
The divergence of the parallel correlation length and wetting film thickness (see inset) obtained using the numerical RG for tricritical wetting, with $\omega=1/4$ (red squares), and critical wetting, with $\omega=0.8$ (green circles), allowing for the Casimir term in the binding potential. The continuous lines correspond to the predicted asymptotic behaviour $\xi_\parallel\sim (t|\ln t|)^{-4/3}$ (see Eq. (\ref{xi_parallel_RG})) and $\xi_\parallel\sim (t|\ln t|^{0.3})^{-3.7}$ (see Ref. \cite{FH_1985}), respectively. The dashed straight line has tangent $\nu_\parallel^{\rm{eff}}=2$ illustrating the effective value of the critical exponent for thinner wetting films.
\label{fig2}}
\end{figure}

These features, together with the onset of asymptotic criticality, can be illustrated most simply by setting $-g=\kappa$ and using the full diagrammatic structure of the functionals $W_{MF}$ and $W_C$ to construct $H_I[\ell]$. When the interface is non-planar the leading order MF attraction and Casimir repulsion remain local, while the non-local MF repulsion vanishes. Consequently, the interfacial Hamiltonian is local, $H_I[\ell]= \int d\mathbf{x} \left[\frac{\Sigma(\ell)}{2} (\nabla\ell)^2+w(\ell)\right]$, containing a Casimir modified binding potential
\begin{equation}
w(\ell)\approx -t \gamma \textrm{e}^{-\kappa\ell}+\frac{\omega \Sigma}{8 }\frac{\textrm{e}^{-2\kappa\ell}}{(\kappa \ell)^2}.
\end{equation}
Here $\Sigma(\ell)= \Sigma + w(\ell)$ is a position dependent stiffness which may be fully accounted for in the RG analysis, although it plays no significant role. Note, that for $0<\omega<1/2$ this potential models tricriticality while, strictly speaking, it models critical wetting for $\omega>1/2$ since the phase boundary is shifted (although the critical behaviour is identical to that for tricritical wetting). In Fig.~\ref{fig2} we show the growth of the parallel correlation length, obtained using the highly accurate numerical, non-linear, RG \cite{FLN, LF_1987} for two values of $\omega$, the larger value corresponding to that pertinent to the 3D Ising model \cite{EHP_1992,FW_1992}. For $\omega=1/4$ there is excellent agreement with the predictions Eqs. (\ref{new_16}) and (\ref{xi_parallel_RG}) over all lengthscales. However, for $\omega=0.8$ the asymptotic regime, where $\nu_\parallel \approx 3.7$, is only reached when $\xi_\parallel$ is mesoscopically large. For thinner wetting layers, for which $10^2 \lesssim \kappa \xi_\parallel \lesssim 10^3$, we find an effective exponent $\nu^{\rm{eff}}_\parallel \approx 2$, close to the value $\nu^{\rm{eff}}_\parallel = 1.8 \pm 0.1$ measured in the Ising model simulations, which corresponded precisely to this range of lengthscales \cite{BB_2013}.

In summary, in this paper we have pointed out that previous theories of 3D short-ranged wetting have missed a thermal Casimir, or entropic, contribution to the binding potential, which we have determined exactly for the LGW model within the DP approximation. This decays exponentially, similar to the MF contribution and is qualitatively different for first-order and critical wetting. Its presence changes the interpretation of thermal fluctuation effects at wetting transitions which arise both from it and from capillary-wave-like interfacial wandering. Both are missing in MF descriptions. The Casimir term strongly affects first-order, critical, and tricritical wetting, where it alters the exponents in all dimensions, including the non-universality in 3D when we allow for interfacial fluctuations. Our central conclusions remain valid beyond the present DP approximation, at least for thick wetting layers. An entropic contribution will be present for other systems and we anticipate it will be similar when there are short-ranged fluid-fluid but long-ranged wall-fluid forces. This is also missing in MF treatments of wetting and may influence surface phase behaviour in the vicinity of the critical point \cite{ESW_2019}. 

\begin{acknowledgments}
J.M.R.-E. acknowledges financial support from Junta de Andaluc\'{\i}a through grants US-1380729 and P20\_00816, cofunded by EU FEDER. A.S. acknowledges LPTMC, Sorbonne Universit\'e for a research stay.
\end{acknowledgments}

\bibliographystyle{apsrev}
\bibliography{bibliography}

\widetext

\newpage
\appendix

\section{Supplementary Material}
Here we outline the derivation of the Casimir contribution using the boundary integral method. This is based on the solution of the Green function $K_\psi (\mathbf{r},\mathbf{r}')$ in the wetting layer satisfying
\begin{equation}
\label{ }
(-\nabla^2+\kappa^2)K_\psi=2\kappa \delta(\mathbf{r}-\mathbf{r}')
\end{equation}
together with the boundary condition on $\mathcal{S}_\psi$ \cite{PRBRE_2006,RESPG_2018},
\begin{equation}
\label{ }
- \frac{1}{g} \hat{\textbf{n}}(\textbf{s}_{\psi}) \cdot \nabla K_{\psi}(\textbf{s},\textbf{s}_{\psi}) = K_{\psi}(\textbf{s},\textbf{s}_{\psi}) \, .
\end{equation}
First we write
\begin{equation}
\label{ }
H[m]=H[m_\Xi]+\Delta H[\delta m] \, ,
\end{equation}
where $\delta m(\mathbf{r})=m-m_\Xi$. Here, within the wetting layer
\begin{equation}
\Delta H[ \delta m] =  \int d\mathbf{r} \left(\frac{1}{2} (\boldsymbol\nabla \delta m)^2 + \frac{\kappa^{2}}{2} (\delta m)^{2} \right) - \frac{g}{2} \int_{\mathcal{S}_\psi} d\mathbf{s} \, (\delta m)^{2} \, ,
\label{HLGW}
\end{equation}
is a Gaussian Hamiltonian with a surface term but only containing the enhancement $g$. The trace over $\delta m$ identifies 
\begin{equation}
\label{oneloop}
\beta W_{C} = \frac{1}{2} \textrm{tr} \ln [K_\psi (K_\psi^\infty)^{-1}] \, ,
\end{equation}
where the superscript corresponds to the unbound interface. The Green function within the wetting layer can be determined using a rigorous perturbative procedure by expanding about the free-space solution $K(\mathbf{r},\mathbf{r}')$, which is a simple Yukawa function - see \cite{PRBRE_2006,RESPG_2018,jointpublication},
\begin{equation}
K(\mathbf{r},\mathbf{r}') = \frac{\kappa}{2\pi} \frac{\textrm{e}^{-\kappa|\mathbf{r}-\mathbf{r}^{\prime}|}}{|\mathbf{r}-\mathbf{r}^{\prime}|} \, .
\end{equation}
Combining the perturbative expansion of $K_{\psi}$ with the exact one-loop result (\ref{oneloop}) gives
\begin{equation}
\beta W_{C}[\ell,\psi]=\frac 1 2\sum_{n=1}^\infty (\Omega_{C})^n_n \, ,
\label{wcasimir}
\end{equation}
where
\begin{equation}
\label{ }
(\Omega_C)_n^n \equiv \frac{1}{n} \textrm{tr} (\mathcal{N}^n) \, ,
\end{equation}
and the operator $\mathcal{N}$ is
\begin{eqnarray}
\mathcal{N}(\mathbf{s},\mathbf{s}') && =\int_{\mathcal{S}_\psi}d\mathbf{s}_1\int_{\mathcal{S}_\psi}d\mathbf{s}_2 \int_{\mathcal{S}_\ell}d\mathbf{s}_3 \Big(K(\mathbf{s},\mathbf{s}_1)\label{defn} + \frac{1}{g}\mathbf{n}(\mathbf{s}_1)\mathbf{\cdot}\boldsymbol\nabla_{\mathbf{s}_1}K(\mathbf{s},\mathbf{s}_1)\Big)X(\mathbf{s}_1,\mathbf{s}_2)K(\mathbf{s}_2,\mathbf{s}_3)K^{-1} (\mathbf{s}_3,\mathbf{s}') \, .
\nonumber\end{eqnarray}
In this expression, the inverse $K^{-1}(\mathbf{s},\mathbf{s}')$ is only required on $\mathcal{S}_\ell$ while $X(\mathbf{s},\mathbf{s}')$, which is the inverse of
\begin{equation}
\label{ }
K(\mathbf{s},\mathbf{s}') + (1/g)\mathbf{n}(\mathbf{s})\mathbf{\cdot} \boldsymbol\nabla_{\mathbf{s}}K(\mathbf{s},\mathbf{s}') - (\kappa/g)\delta(\mathbf{s}-\mathbf{s}') \, ,
\end{equation}
(with ${\bf{n}}$ the normal), is only required on 
$\mathcal{S}_\psi$. 
Each can be expressed as a series of convolutions involving
\begin{equation}
\label{ }
U(\mathbf{s},\mathbf{s}')\equiv K(\mathbf{s},\mathbf{s}')-\delta(\mathbf{s}-\mathbf{s}') \, ,
\end{equation}
and
$\mathbf{n}(\mathbf{s})\cdot 
\boldsymbol\nabla_{\mathbf{s}}K(\mathbf{s},
\mathbf{s}')$. Here we
 focus on the leading terms in
 Eq.~(\ref{wcasimir}) for large distance between 
$\mathcal{S}_\ell$ and $\mathcal{S}_\psi$ and 
large radii of curvature as relevant to discussions of
 wetting.

We now recast the expression (\ref{defn}) diagrammatically using the same notation as in \cite{RESPG_2018}
\begin{equation}
\label{defdiagrams}
K(\mathbf{s},\mathbf{s}')\equiv\fig{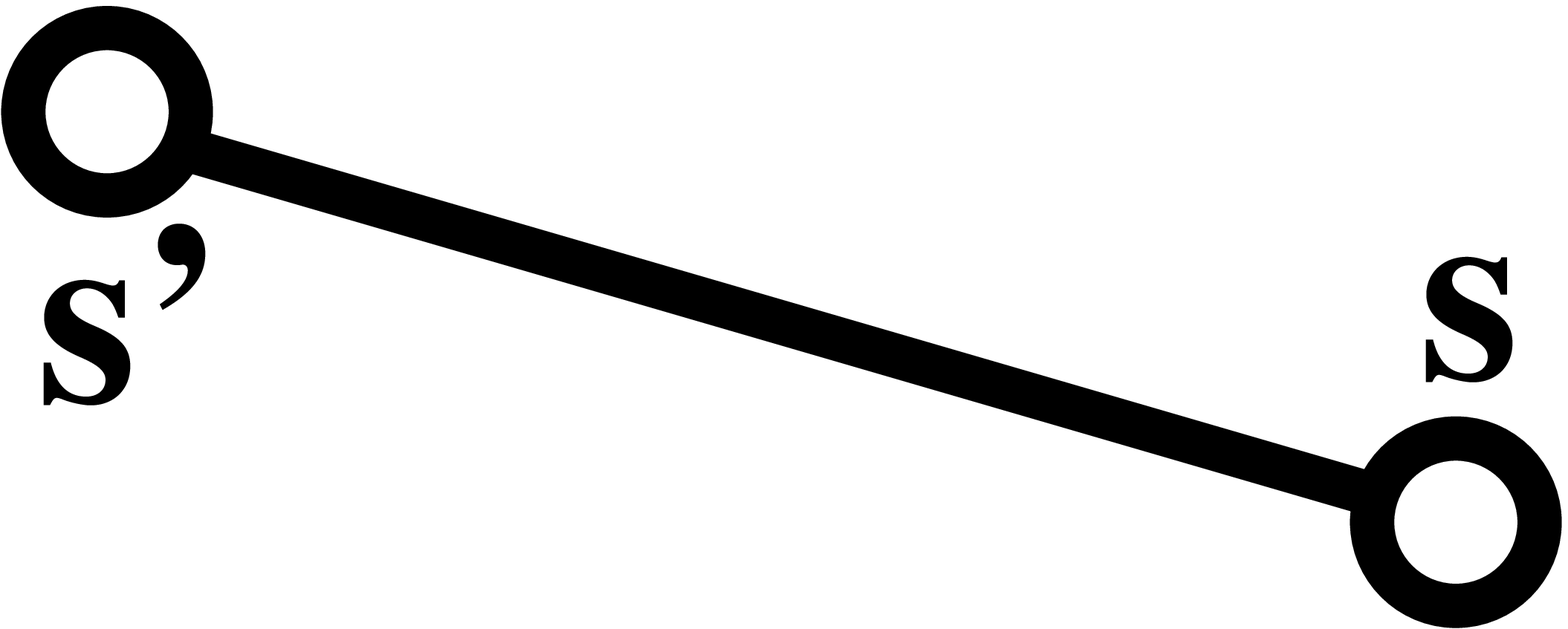} \qquad, \qquad U(\mathbf{s},\mathbf{s}')\equiv\fig{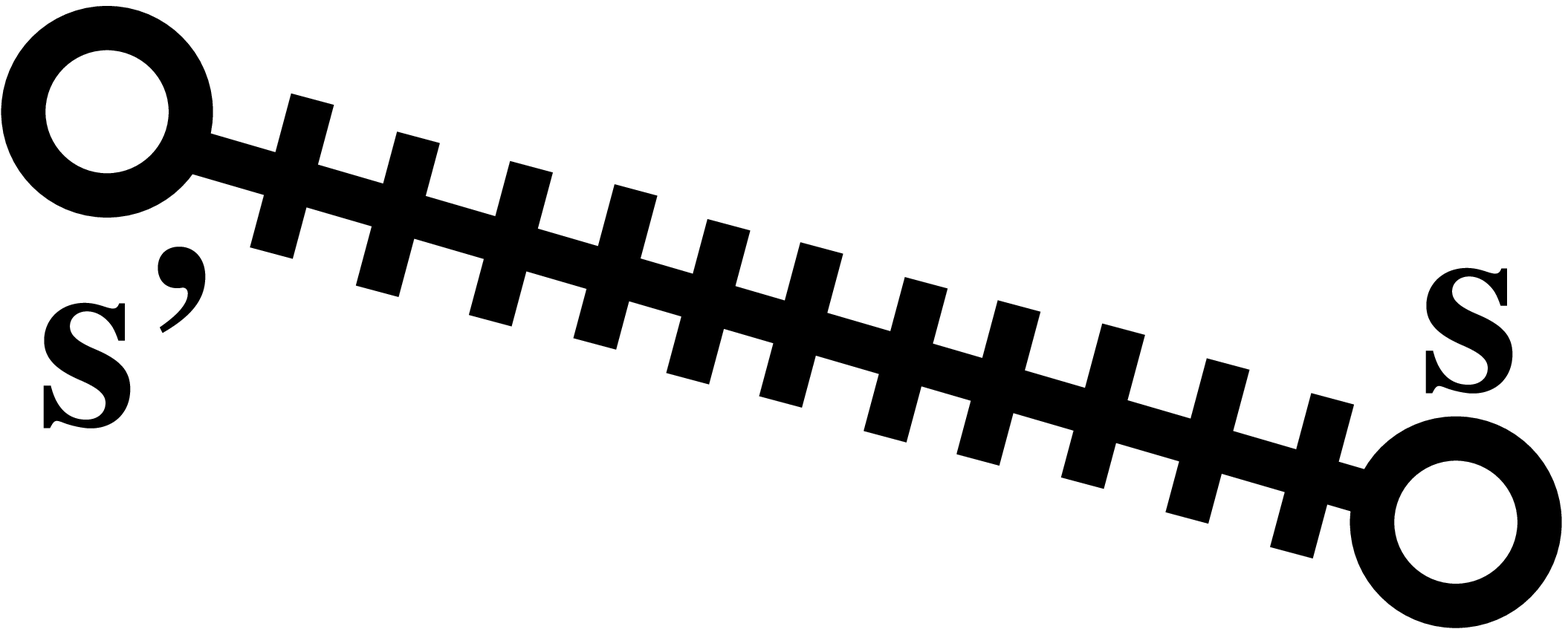} \qquad, \qquad \frac{1}{\kappa}\mathbf{n}(\mathbf{s})\cdot\boldsymbol\nabla_{\mathbf{s}} K(\mathbf{s},\mathbf{s}') 
\equiv \fig{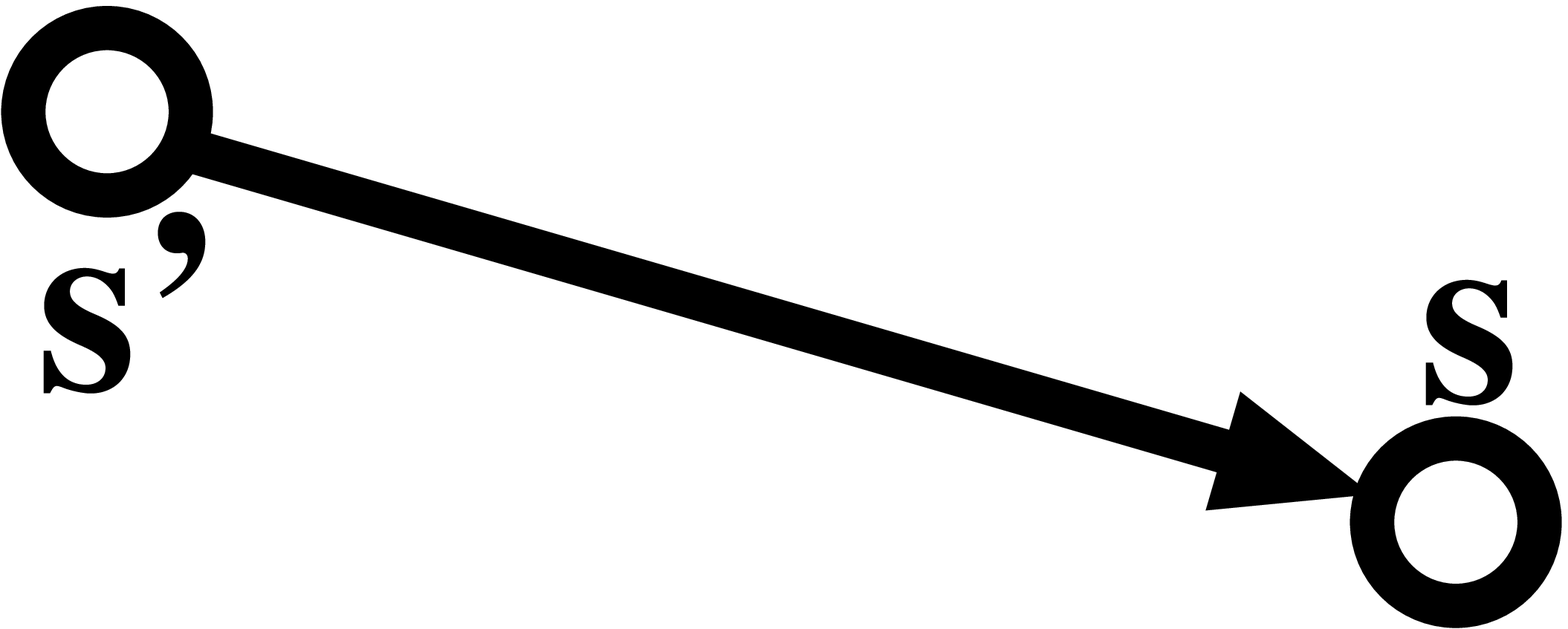}
\end{equation}
The integrals in $\mathcal{N}$ can also be represented diagrammatically. Each connect the interface (top wavy line) with the wall (bottom wavy line). We define
\begin{eqnarray}
\fig{fig6}=\int_{\mathcal{S}_\psi} d\mathbf{s}_1\left(K+\frac{1}{g}\mathbf{n}\mathbf{\cdot}\boldsymbol\nabla K\right)X
 \label{defdiagrams2}
\end{eqnarray}
where $K=K(\mathbf{s},\mathbf{s}_1)$ and $X=X(\mathbf{s}_1,\mathbf{s}')$. This new diagram can be written $\alpha\mathcal{I}+(1-\alpha)\mathcal{J}$ where $\alpha=g/(g-\kappa)$,
\begin{equation}
\mathcal{I}=\fig{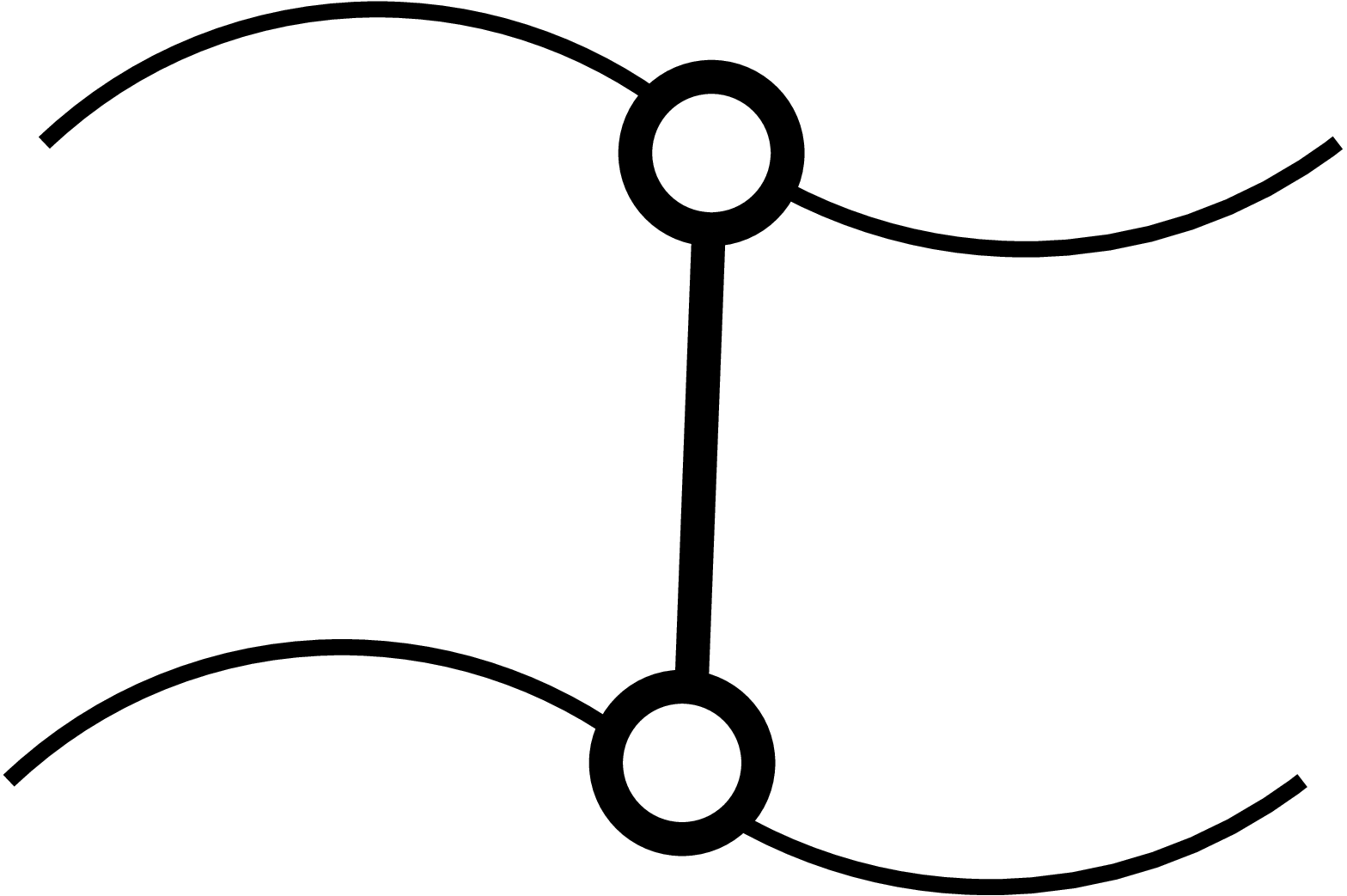}-\alpha \fig{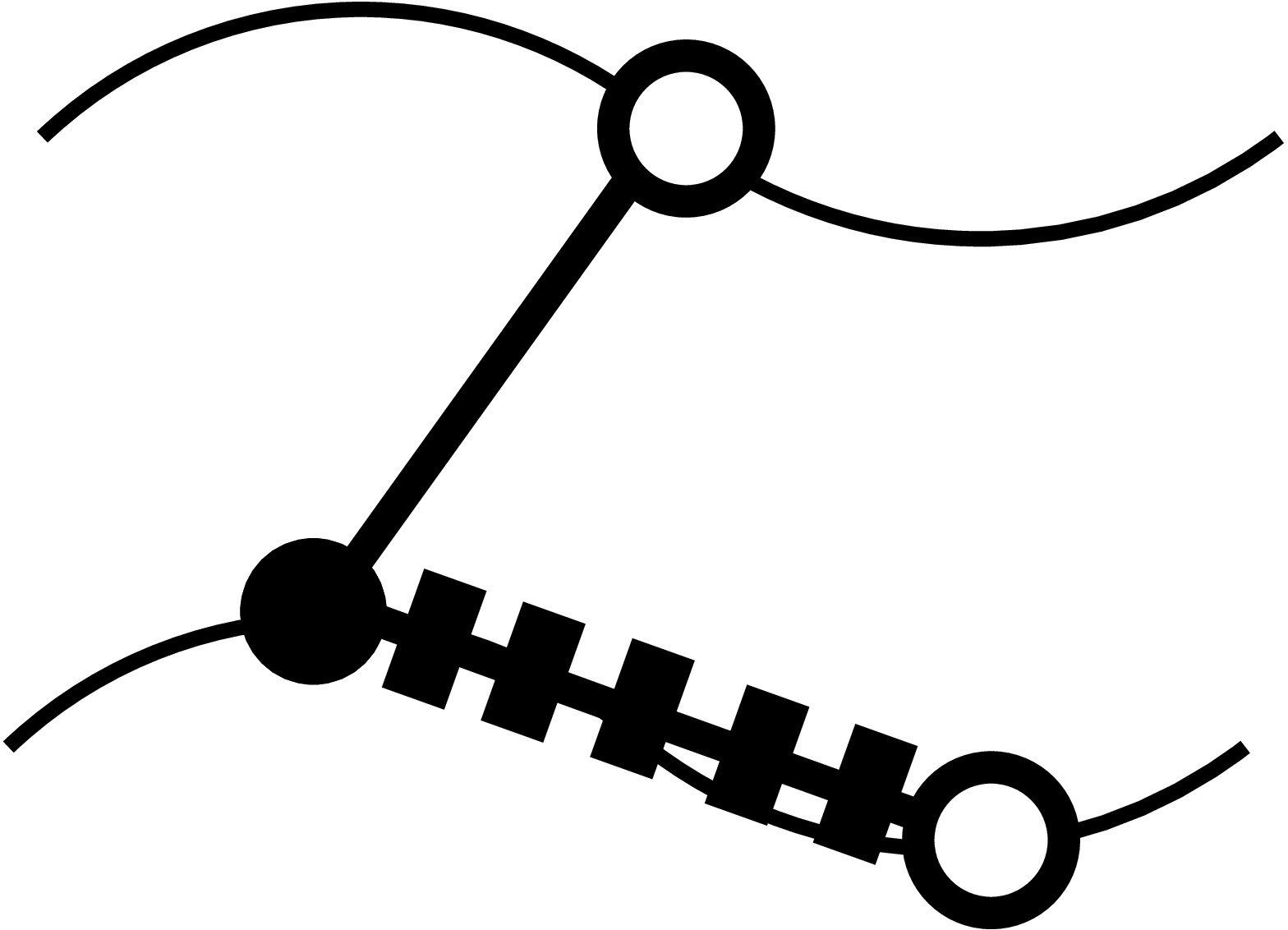} + \alpha^2 \fig{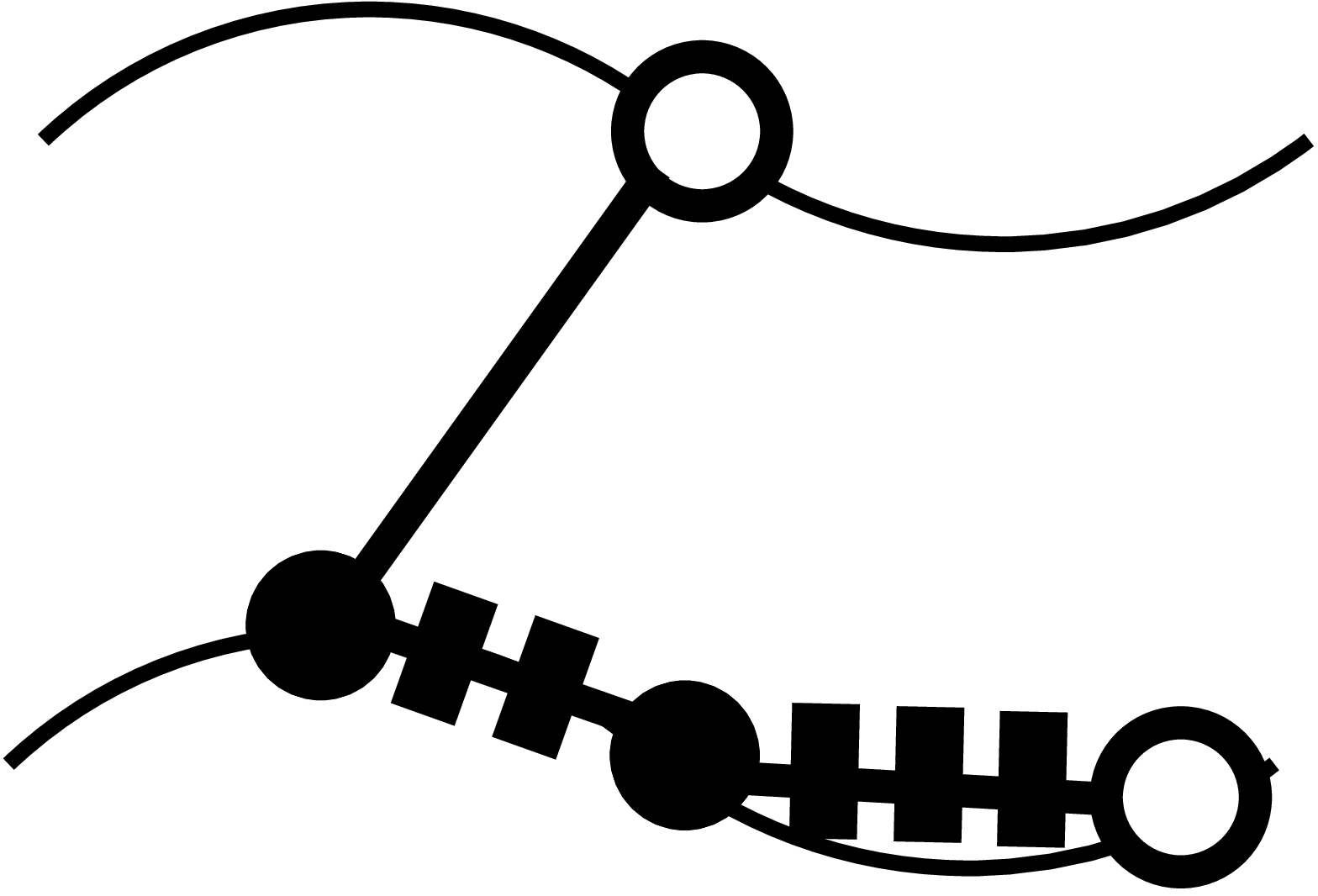}+\ldots 
\label{defdiagrams2-1}
\end{equation}
and
\begin{equation}
\mathcal{J}=\fig{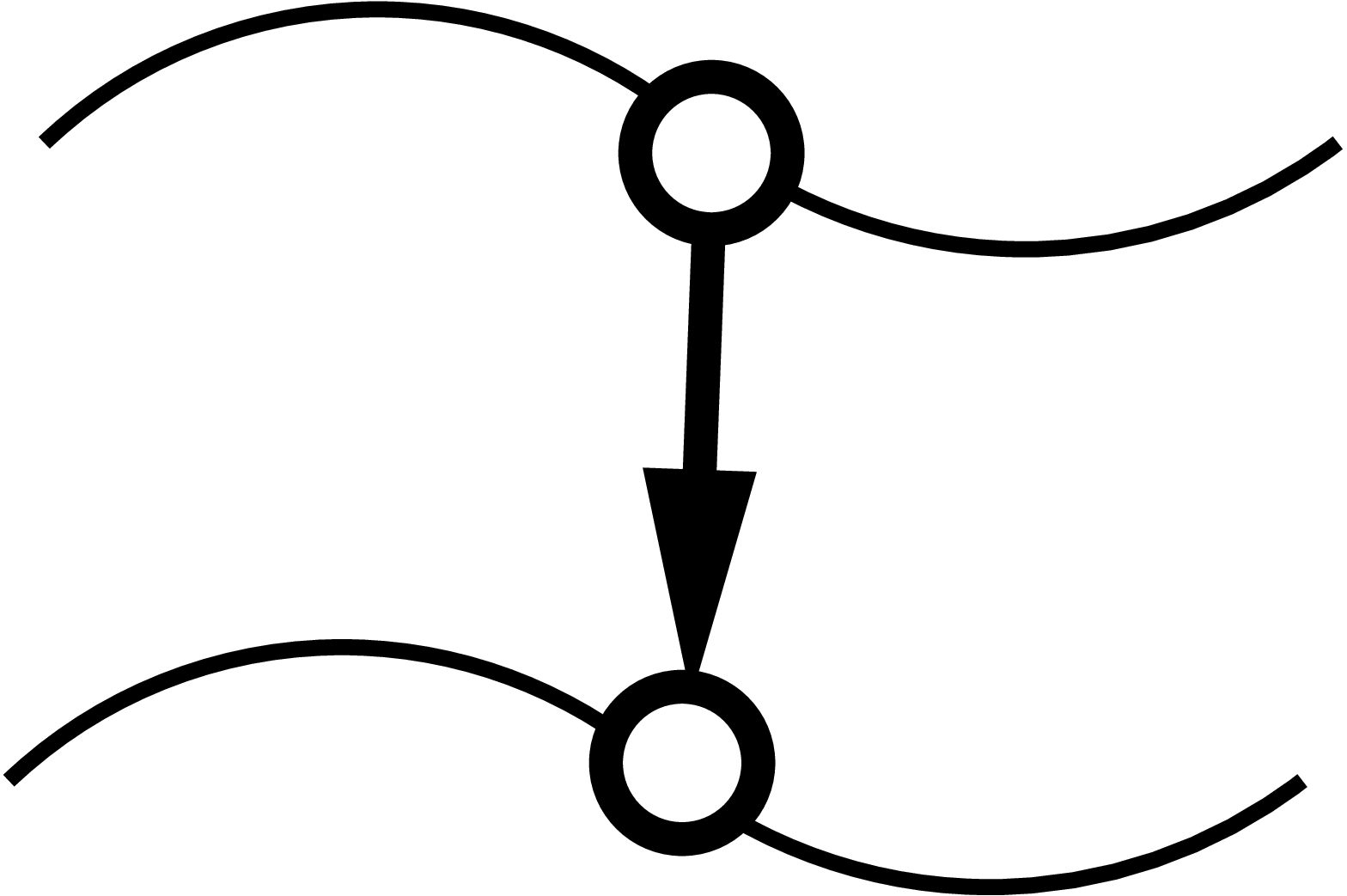}-\alpha \fig{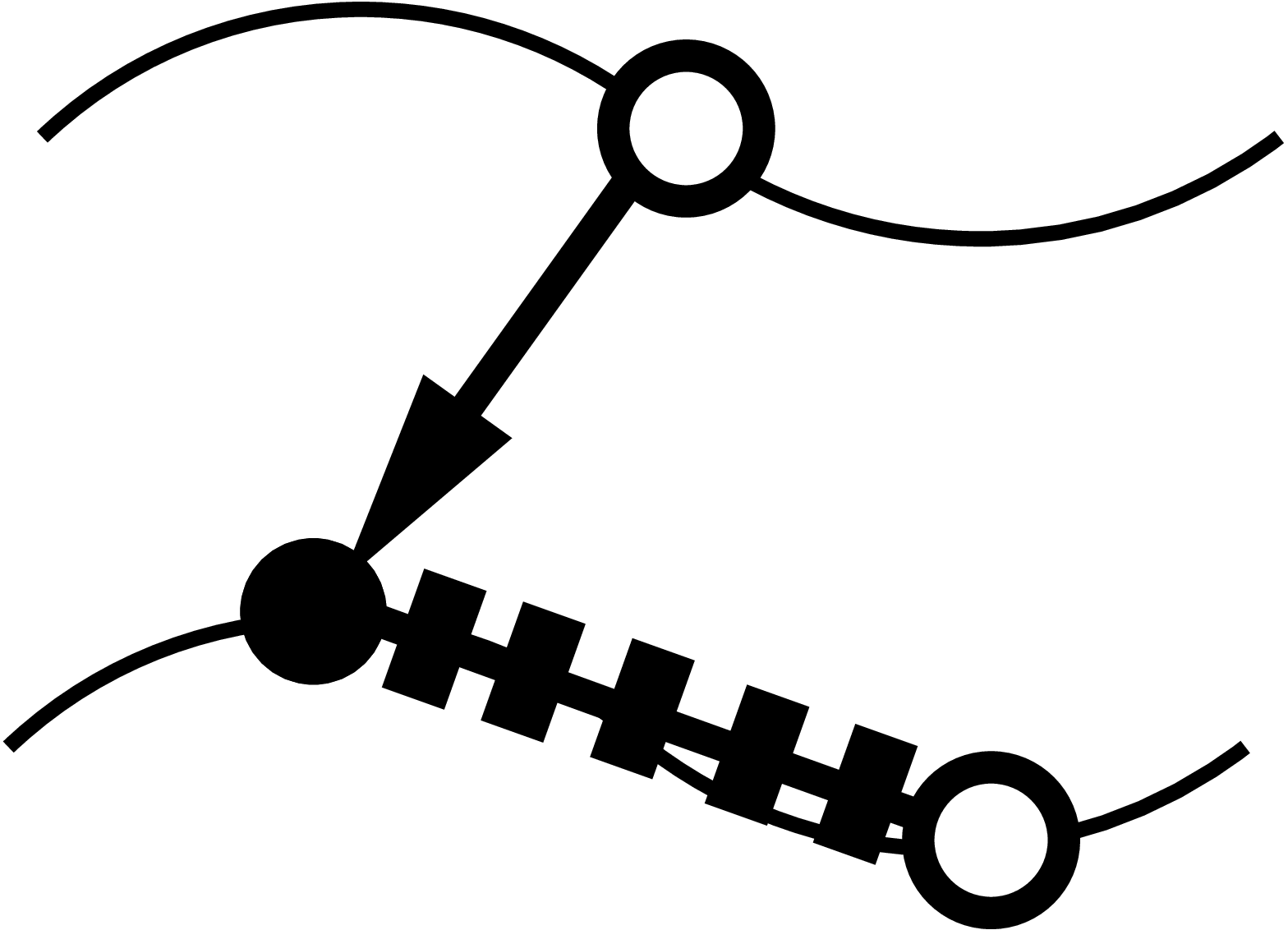} + \alpha^2 \fig{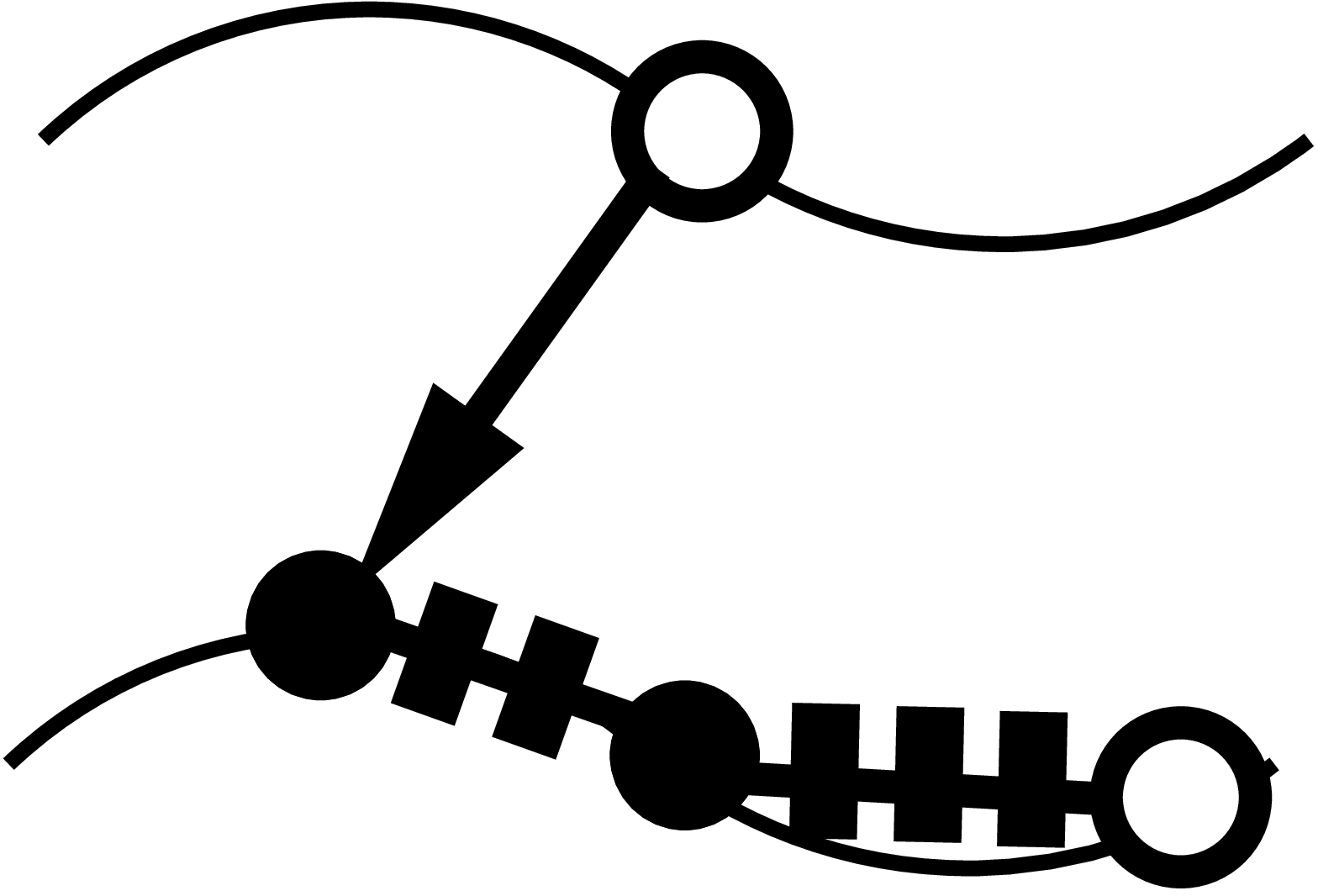}+\ldots\ , 
\label{defdiagrams2-2}
\end{equation}
where the black dot means integration over the surface. These resum to the explicit algebraic expression
 \begin{eqnarray}
  \fig{fig6} 
 \approx \frac{1}{2\pi} \int_0^\infty dq\ q \frac{g+\kappa_q}{g-\kappa_q} J_0(q\rho)\exp(-\kappa_q \ell), 
  \label{defdiagrams2bisbis} \end{eqnarray}
 where $\rho$ and $\ell$ are, respectively, the transverse and normal coordinates of $\mathbf{s}-\mathbf{s}'$, $J_0(z)$ is the 
Bessel function of the first kind and zero order and 
$\kappa_q\equiv \sqrt{\kappa^2+q^2}$. 
The last integral in $\mathcal{N}$ can be expressed
\begin{equation}
\int_{\mathcal{S}_\ell} d\mathbf{s}_1 K(\mathbf{s}',\mathbf{s}_1)K^{-1}(\mathbf{s}_1,\mathbf{s})
 = -\fig{fig13} ,\label{defdiagrams3}
\end{equation}
where the arrow diagram is the same as that appearing 
in the dictionary (\ref{defdiagrams}), given 
explicitly by
\begin{equation}
 -\fig{fig13} =  \mathbf{n}(\mathbf{s}) \mathbf{\cdot} \frac{\mathbf{s}-\mathbf{s}^{\prime}}{|\mathbf{s}-\mathbf{s}^{\prime}|^2}\left(1+ \frac{1}{\kappa|\mathbf{s}-\mathbf{s}^{\prime}|}\right)\textrm{e}^{-\kappa|\mathbf{s}-\mathbf{s}^{\prime}|}.
\label{defdiagrams3bisbis}
\end{equation}
This leads to the diagrammatic expansion
\begin{equation}
\beta W_{C}[\ell,\psi] = \frac{1}{2}\left(\fig{fig14}-\frac{1}{2}\fig{fig15}+\ldots\right),
\label{wcasimirdiagrammatic}
\end{equation} 
which is our central result.

\end{document}